\documentclass[12pt]{article}

\usepackage{amsmath}
\usepackage{graphicx}

\title{ Vacuum  phase transition at nonzero baryon density.}
\author{ Yu.A.Simonov, M.A.Trusov\\
State Research
Center\\Institute of Theoretical and Experimental Physics, \\
Moscow, 117218 Russia}

\date{}

\newcommand{\beq}{\begin{eqnarray}}
 \newcommand{\eeq}{\end{eqnarray}}
\newcommand{\be}{\begin{equation}}
 \newcommand{\ee}{\end{equation}}

 \def\la{\mathrel{\mathpalette\fun <}}

\def\fun#1#2{\lower3.6pt\vbox{\baselineskip0pt\lineskip.9pt
\ialign{$\mathsurround=0pt#1\hfil ##\hfil$\crcr#2\crcr\sim\crcr}}}

\newcommand{{\SD}}{\rm SD}

\newcommand{\lan}{\langle}
\newcommand{\ran}{\rangle}

\begin{document}
\maketitle
\begin{abstract}

It is argued that the dominant  contribution to the interaction of
quark gluon plasma at moderate $T\geq T_c$ is given by the
nonperturbative vacuum field correlators.
 Basing on that nonperturbative equation of
state of quark-gluon plasma is computed and in the lowest
approximation expressed in terms of absolute values of Polyakov
lines for quarks and gluons $L_{fund} (T);
L_{adj}(T)=(L_{fund})^{9/4}$known  from  lattice and analytic
calculations.
 Phase transition at any $\mu$ is described as a transition due to vanishing of
one of correlators, $ D^E(x)$,
 which implies the change of gluonic condensate $\Delta G_2$.
  Resulting transition temperature $T_c(\mu)$ is
 calculated in terms of  $\Delta$$G_2$ and $L_{fund}(T_c)$.

 The phase curve $T_c(\mu)$ is in  good agreement with
 lattice data. In particular $T_c(0)=0.27; 0.19; 0.17$ GeV for
 $n_f=0,2,3$ and fixed $\Delta G_2=0.0035$ GeV$^4$.

\end{abstract}

\section{Introduction} QCD at nonzero  temperature $T$ and quark
chemical potential $\mu$, and specifically phase transitions are
important topics both for theory, and  experiment \cite{1,2} and
are carefully studied on the lattice  (for recent reviews see
\cite{3,4}).

Since these phenomena cannot  be explained in perturbative QCD,
any analytic approach should be based on nonperturbative methods,
and as such solvable models, like Nambu-Jona-Lasinio model, have
been used mostly. However the models used do not contain
confinement and treat phase transition from the chiral symmetry
(CS)  point of view. Therefore it is interesting  to  look at the
Equation of State (EoS) and phase transition in the framework of a
nonperturbative (NP) method, based on vacuum fields and
implementing confinement.

In this letter we are using such tool, which is called the Field
Correlator Method (FCM) \cite{5}. In FCM all NP information about
QCD vacuum is contained in field correlators, and the Casimir
scaling, now proved on the lattice also for $T\geq T_c$ \cite{6},
allows ( with $\sim 1\%$ accuracy) to  consider only four
quadratic correlators for $T,\mu>0$ of colorelectric
(colormagnetic) field strengths, $D^E(x), D_1^E(x), (D^H(x),
D_1^H(x))$. Here $D^E, D^E_1$ are defined as vacuum averages of
colorelectric field $E_j $,
$$
 {g^2 } \langle  \hat tr_f [ E_i (x) \Phi(x,y) E_k (y)
\Phi(y,x) ] \rangle  =$$ $$ = \delta_{ik} \left[ {D}^E + {D}_1^E +
u_4^2 {\partial {D}_1^E \over \partial u_4^2} \right] + u_i u_k
{\partial {D}_1^E \over
\partial \vec{u}^2}$$
and similarly for $D^H, D_1^H$ with colormagnetic fields $H_i$
instead of $E_i$, where the sign of the last term is reversed, and
$\Phi(x,y) =P\exp ig \int^x_y A_\mu dz_\mu$ is parallel
transporter (Schwinger line). Note that only $D^E$ contributes to
the string tension $\sigma^E $ implying confinement, and $D^H$ to
the spatial string tension, $\sigma^H\equiv \sigma_s$, while
$D^E_1, D^H_1$ contain both perturbative and nonperturbative
parts.

 The correlators were calculated on
the lattice \cite{7,8} and also analytically \cite{9}.

Most calculations with FCM at $T=\mu=0$ contain as input only
string  tension $\sigma$, strong coupling $\alpha_s(q)$ and
current (pole) quark masses, and results for hadronic spectra are
in a good agreement with available experimental and lattice data,
see last ref. in \cite{5} and \cite{10} for reviews.

In \cite{11} the FCM was extended to the case of nonzero $T$, and
in \cite{12} the phase transition at $\mu=0$ was described as a
Vacuum Dominated (VD) transition from  the  confining vacuum (all
$D^{(i)}, D_1^{(i)}, i=E,H$ are nonzero, and $D^E=D^H,
D_1^E=D_!^H$ as for $T=0$) to the deconfined state (only $D^E$ and
hence $\sigma^E =\frac12 \int D^E (x) d^2 x$ are zero). This
picture was supported by lattice measurements \cite{7,8} where
indeed $D^E$ was shown to vanish at $T=T_c$, while $D^E_1$ and
colormagnetic correlators stay intact. This fact supports the
conjecture \cite{13}  that $D_1^E $ is  nonzero at $T>T_c$ and
strong enough to support bound states of quarks and gluons. Bound
states of $q\bar q$ were indeed found on the lattice years later
(see \cite{4} for a review) and in \cite{14, 15} the connection of
those to the correlator $D_1^E$ was quantitatively established.

Moreover in \cite{14} it was shown, that  a new  static potential
$V_1(r,T)$ is provided  by $D_1^E$ at $T\geq T_c$ and  the NP part
of  it yields the modulus of the  renormalized Polyakov line,
$L_{fund} =exp \left(-\frac{V_1(\infty, T)}{2T}\right)$.

The physical picture of the  VD deconfining phase transition given
in \cite{12}, has allowed to get  the  (simplified) estimate of
the transition temperature $T_c$ at $\mu=0$  in reasonable
agreement with lattice data for $n_f =0, 2,3$ the only input being
the standard value of the gluonic condensate $G_2$, known from the
ITEP sum rule method \cite{16}. It is the purpose of the  present
letter to extend the method to the case of nonzero $\mu$ and
nonzero interaction with vacuum $(V_1\neq 0)$ and to   find
$T_c(\mu)$,  for $\mu>0$.

To this end we are using the EoS found recently in \cite{17},
where pressure in the lowest approximation is expressed through
the only dynamical quantity -- $L_{fund} (T)$ calculated via the
correlator $D_1^E(x)$ or obtained from lattice data. This factor
takes into account interaction of the quark with the NP vacuum  in
the form of $L_{fund}(T)$  and analogous factor for gluons is
$L_{adj} (T) = (L_{fund})^{9/4}$. The interaction between quarks
and gluons is considered as a next step, and is argued  to
possibly contribute in a narrow temperature region near $T_c$
\cite{17,18}. This simple picture of VD dynamics  with the only
input $L_{fund}(T)$, taken from lattice or analytic calculation
suggested in \cite{17}, which may be called Vacuum Dominance Model
(VDM), is adopted below and is shown to produce surprisingly
reasonable results, being in good agreement with available lattice
data for $\mu=0$ or nonzero $\mu$, where these data are reliable.

\section{Nonperturbative EoS for $\mu>0$}

The main idea of the  VDM  discussed below is that the most
important part of quark and gluon dynamics in the strongly
interacting plasma is the interaction of each individual quark or
gluon with vacuum fields. This interaction is derived from field
correlators and is rigorously proved to be embodied in   factors,
which happen to coincide with the modulus of Polyakov
loop\footnote{We neglect in the approximation the difference
between $L_{fund}$ expressed via $V_1(\infty, T)$ and
$L_{fund}^{lat}$ where the role of $V_1(T)$ is played \cite{14,17}
by singlet $Q\bar Q$ free energy $F^1_{Q\bar Q} (\infty, T)$. The
latter quantity contains all excited states, so that $V_1(T)\geq
F^1_{Q\bar Q} (\infty, T)$.}, \be L_{fund} =\exp (-\frac{V_1
(T)+2V_D}{2T}),~~L_{adj}=\exp (-\frac94
\frac{V_1(T)+2V_D}{2T})\label{1}\ee where $V_1(T) \equiv V_1
(\infty, T),$  $V_D\equiv V_D(r^*,T)$ and $V_1(r,T), V_D$ found in
\cite{14} to be  ($\beta\equiv 1/T)$ \be V_1(r,T)= \int^\beta_0
d\nu (1-\nu T) \int^r_0 \xi d \xi
D^E_1(\sqrt{\xi^2+\nu^2}),\label{2}\ee \be V_D(r,T)= 2\int^\beta_0
d\nu (1-\nu T) \int^r_0 (r-\xi) d \xi
D^E(\sqrt{\xi^2+\nu^2}).\label{3}\ee

In $V_D, $  Eq.(\ref{1}) $r^*$ is the average size of the
heavy-light $Q\bar q$ or its adjoint equivalent system; for
$T>T_c$ one has $D^E=V_D=0$. For $T<T_c$ one has
$r^*_{fund}=\infty$ for $n_f=0$, yielding $L_{fund}=0,$ however
$r^*_{adj}\approx 0.4$ fm for any $n_f$ and gives nonzero $L_{adj}
(T<T_c)$. The form (\ref{1})-(\ref{3}) is in good agreement with
lattice data \cite{6} and also explains why $L_{fund}$ is a good
order parameter (however approximate for $n_f\neq 0$). Note, that
only NP parts $D_1^E, D^E$ enter in (\ref{2}), (\ref{3}), see
first ref. in \cite{9} for discussion of separation of these
parts; renormalization procedure is discussed in \cite{14,15}.

In the lowest NP approximation one neglects pair, triple etc.
interactions between quarks and gluons (which are important for
$T_c\leq T\leq 1.2 T_c$ where density is low and screening by
medium is  not yet operating, see \cite{17,18})) and derives the
following EoS (this approximation is called in \cite{17} the
Single Line Approximation (SLA)) \be p_q\equiv{\frac{
P^{SLA}_{q}}{T^4}} = \frac{4N_cn_f}{{\pi}^2}
\sum^\infty_{n=1}\frac{ (-1)^{n+1}}{n^4}  L^{n}_{fund}
\varphi^{(n)}_q cosh\frac{\mu n}{T}\label{4}\ee

\be p_{gl} \equiv \frac{P_{gl}^{SLA}}{T^4}
=\frac{2(N^2_c-1)}{\pi^2} \sum^\infty_{n=1} \frac{1}{n^4}
L_{adj}^n \label{5}\ee with \be \varphi^{(n)}_q (T)
=\frac{n^2m^2_q}{2T^2} K_2 \left(\frac{m_q n}{T}\right)\approx
1-\frac{1}{4} \left(\frac{n m_q}{T}\right)^2+...\label{6}\ee

In (\ref{4}), (\ref{5}) it was assumed that $T\la
\frac{1}{\lambda}\cong 1$ GeV, where $\lambda$ is the vacuum
correlation length, e.g. $D_1^{(E)} (x) \sim e^{-|x|/\lambda}$,
hence powers of $L^n_i$,  see \cite{17} for details.

With few percent accuracy one can replace the sum in (\ref{5}) by
the first term, $n=1$, and this form will be used below for
$p_{gl}$, while for $p_q$ this replacement is not valid for large
$\frac{\mu}{T}$, and one can use instead the form equivalent to
(\ref{4}), \be p_q=\frac{n_f}{\pi^2} \left[ \Phi_\nu \left(
\frac{\mu-\frac{V_1}{2}}{T}\right)+ \Phi_\nu \left(-
\frac{\mu+\frac{V_1}{2}}{T}\right)\right]\label{7}\ee where
$\nu=m_q/T$ and \be \Phi_\nu (a) =\int^\infty_0  \frac{z^4
dz}{\sqrt{z^2+\nu^2}}\frac{1}{(e^{\sqrt{z^2+\nu^2}-a}+1)}.\label{8}\ee

Eqs. (\ref{7}), (\ref{5}) define $p_q, p_{gl}$ for all $T, \mu$
and $m_q$, which is the current (pole) quark mass at the scale of
the order of $T$.

To draw $p_q, p_{gl}$ and $p\equiv  p_q+p_{gl}$ as functions of
$T, \mu$ one needs explicit form of $V_1(T)$. This was obtained
analytically and discussed in \cite{14}; another form was found
from  $D_1^E(x)$ measured on the lattice in \cite{8,9} and is
given in \cite{15}.

Also from lattice   correlator  studies \cite{8,9}, \cite{14}
$V_1(T=T_c)$ is (with $\sim 10\% $ accuracy) 0.5 GeV and is
decreasing with the growth of $T$ (cf. Fig. 2 of \cite{14} and
Fig. 1 of \cite{15}). This behaviour is similar to that found
repeatedly on the lattice  direct measurement of $F^{(1)}_\infty$,
see e.g. Fig.2 of \cite{19} where $F_\infty^{(1)} =V_1(T)$ is
given for $n_f =0,2,3$. In what follows we shall exploit the
latter curves parametrizing them for $T\geq T_c$ and all $n_f$  as
\be V_1(T) =\frac{0.175~{\rm GeV}}{1.35\left(\frac{T}{T_c}\right)
-1},~~ V_1(T_c) \approx 0.5 {~\rm GeV}.\label{9}\ee For $\mu>0$
one  can expect a $\mu$-dependence of $V_1$, however it should be
weak for values of $\mu$ much smaller than the scale of change of
vacuum fields. The latter scale can  be identified with the
dilaton mass $m_d$, which is of the  order of the lowest glueball
mass, i.e. $\sim  1.5$ GeV $\equiv  m_d$. Hence one can expect,
that $V_1$ in the lowest approximation does not depend on $\mu$.
This is supported by the lattice measurements in \cite{20}, where
for $\frac{T}{T_c}=1.5$ and $\frac{\mu}{T} =0.8$ the values of
$F^{(1)}_\infty$ are almost indistinguishable from the case of
$\frac{\mu}{T}=0$.

To give an illustration of the resulting EoS we draw in Fig.1 the
pressure $p$ for the cases $\mu=0$, $n_f=0,2$. One can see a
reasonable behaviour  for $T_c(n_f) =(0.27; 0.19; 0.17) $ GeV (for
$n_f=0,2,3$ respectively)similar to the  lattice data,  see
\cite{21} for a review and discussion.

\begin{figure}
\includegraphics[width=12cm]{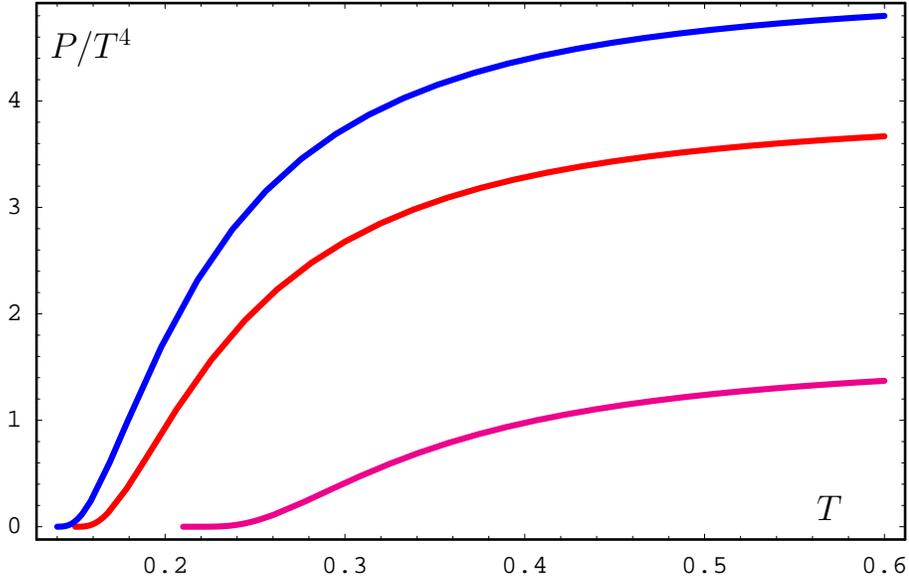}
\caption{Pressure $\frac{P}{T^4}$ from Eq.(\ref{4},\ref{5}) as
function of temperature  $T$ (in GeV) for $ n_f =3,2,0$ (top to
bottom) and $\Delta G_2=0.0034$ GeV$^4$.}
\end{figure}

\section{Phase transition for nonzero $\mu$}

Here we extend the  VD mechanism suggested  in \cite{12} to the
case of nonzero $\mu$ and $V_1$. We assume as was said in
Introduction that the phase transition occurs from the full
confining vacuum with all correlators $D^E, D_1^E, D^H, D^H_1$
present to the deconfined vacuum where $D^E$ vanishes. The basics
of our physical picture is  that all fields (and correlators) do
not change for $\mu, T$ changing in a wide interval, unless $\mu,
T$ become comparable to the dilaton mass $m_d \approx M$(glueball
$0^+)\approx 1.5$ GeV\footnote{Note that $G_2$ does  not depend on
$\mu, n_f$ in the leading order of the $1/N_c$ expansion and one
expects a growth of the magnetic part of $G_2$ at $T>T_{dim.red}
\approx 2 T_c$, $G_2^{mag} \approx O(T^4)$, in the regime of
dimensional reduction. }. Therefore correlators and $\sigma^E,
\sigma^H$ are almost constant till $T=T_c$ and  at $T\geq T_c$ a
new vacuum phase with $D^E=\sigma^E =0$ is realised,which yields
lower thermodynamic potential (higher pressure). Lattice
measurements \cite{7,8} support this picture. The crucial step is
that one should take into account in the free energy $F$ of the
system also the free energy of the vacuum, i.e. vacuum energy
density $\varepsilon_{vac} =\frac{1}{4} \theta_{\mu\mu} =
\frac{\beta(\alpha_s)}{16\alpha_s} \lan (F^a_{\mu\nu})^2\ran=
-\frac{(11-\frac23 n_f)}{32} G_2$, which can be estimated via the
standard gluonic  condensate \cite{16} $G_2\equiv
\frac{\alpha_s}{\pi} \lan (F^a_{\mu\nu})^2\ran\approx 0.012$
GeV$^4$.

Hence for the pressure $P=-F$ one can write in  the phase $I$
(confined) \be P_I =|\varepsilon_{vac}|+ \chi_1(T)\label{10}\ee
where $\chi(T)$ is the hadronic gas pressure, starting with pions,
$\chi_{pion} \cong \frac{\pi^2}{30} T^4$. In the deconfined phase
one can write \be P_{II} = |\varepsilon^{dec}_{vac}| + (p_{gl} +
p_q) T^4\label{11}\ee where $|\varepsilon^{dec}_{vac}|$ is  the
vacuum energy density in the deconfined phase, which is mostly
(apart from $D^E_1(0)\approx 0.2 D^E(0)$ see
 \cite{7} and first ref.
of \cite{9}) colormagnetic energy density and by the same
reasoning as before we take it as for $T=0$, i.e.
$|\varepsilon^{dec}_{vac}| \cong 0.5|\varepsilon_{vac}|$.

Equalizing $P_I$ and $P_{II}$ at  $T=T_c(\mu)$ one obtains the
equation for $T_c$ \be T_c(\mu)
=\left(\frac{\Delta|\varepsilon_{vac}|+\chi(T)}{p_{gl}+p_q}\right)^{1/4}
\label{12}\ee where $\Delta|\varepsilon_{vac}| =
|\varepsilon_{vac} | - | \varepsilon_{vac}^{dec}| \approx
\frac{(11-\frac23 n_f)}{32} \Delta G_2 $;~ $\Delta G_2\approx
\frac12 G_2;$ $p_{gl}$ and $p_q$ are given in (\ref{5}), (\ref{7})
respectively and depend on both $T_c$ and $\mu$.

In this letter we shall consider the simplest case when the
contribution of hadronic gas $\chi_1(T)$ can be  neglected in the
first approximation. Indeed, pionic gas  yields only $\sim 7$\%
correction to the numerator of (\ref{12}) at $T\approx T_c$, and
from \cite{22a} one concludes that $\chi(T_c)\la 0.5 T^4_c$, which
yields a $\la 10\%$ increase of $T_c$ for $G_2\sim 0.01$ GeV.

 From  the expression for $T_c(\mu)$ (\ref{12}) one can find limiting
 behaviour of $T_c(\mu\to 0)$ and $\mu_c(T\to 0)$. For the first
 one can use for $p_q$ and $p_g$ (\ref{4}) and (\ref{5}) and
 expand r.h.s. of (\ref{12}) in ratio $p_g/p_q$ with the result.
 \be T_c=T^{(0)} \left( 1+ \frac{V_1(T_c)}{8T_c} + O\left(
 \left(\frac{V_1(T_c)}{8T_c}\right)^2\right)\right)\label{13}\ee
where the last term yields a 3\% correction, and $T^{(0)}
=\left(\frac{(11-\frac23 n_f) \pi^2\Delta G_2}{32\cdot 12
n_f}\right)^{1/4}$. Solving (\ref{13}) for $T_c$ one has \be T_c
=\frac12 T^{(0)} \left(1+\sqrt{ 1+\frac{\kappa}{T^{(0)}}}\right)
\left(1+\frac{m^2_q}{16 T^2_c}\right)\label{14} \ee with
$\kappa\equiv \frac12 V_1(T_c)$. From (\ref{12}), (\ref{13}) one
can compute expansion $T_c(\mu)$ in powers of $\mu$,
$$ T_c(\mu_B)=T_c(0)  \left(1-C\frac{\mu^2_B}{T^2_c(0)}\right), ~~
\mu_B =3\mu.$$

$$C=\frac{1+
\sqrt{1+\frac{\kappa}{T^{(0)}}}}{144\sqrt{1+\frac{\kappa}{T^{(0)}}}}
=0.0110(3)~~ {\rm for} ~ n_f=2,3,4.$$

One can see, that $C$ practically does not depend on $n_f$ and is
in the same ballpark as  the values  found by lattice
calculations, see \cite{27}, \cite{3} for  reviews and references.

Another end point of the phase curve, $\mu_c(T\to 0)$, is found
from (\ref{12}) when one takes into account asymptotics $\Phi_0 (a
\to \infty) =\frac{a^4}{4} +\frac{\pi^2}{2} a^2 +
\frac{7\pi^4}{60}+...$, which yields (for small $\frac{m_q}{\mu}$)
\be \mu_c (T\to 0) = \frac{V_1(T_c)}{2} + (48)^{1/4} T^{(0)}
\left( 1+ \frac{3m^2_q}{4\mu^2_c}\right)\left( 1-\frac{\pi^2}{2}
\frac{T^2}{\left(\mu_c-\frac{V_1(T_c)}{2}\right)^2}\right)\label{15}
\ee

 The resulting curve $T_c(\mu)$ according to Eq.(\ref{12}) with
 $\chi_1\equiv 0$ is given  in Fig. 2 for$\Delta G_2 =0.00341 $GeV$^4$,
$n_f=2,3$
 and $m_q=0$.

\begin{figure}[htb]
\includegraphics[height=7cm]{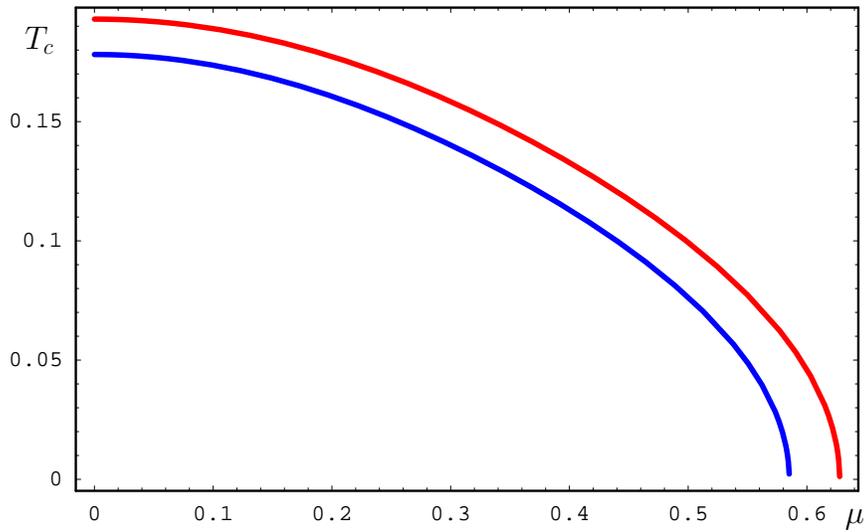}
\caption{The phase transition curve $T_c(\mu)$ from Eq.(\ref{12})
(in GeV) as function of quark chemical potential $\mu$ (in GeV)
for $n_f=2$ (upper curve ) and $ n_f=3$ (lower curve) and $\Delta
G_2=0.0034$ GeV$^4$.}.
\end{figure}

 \section{Discussion of results}

 Our prediction  of $T_c(\mu)$ depends only on two numbers: 1) the value of
gluonic condensate $ \Delta G_2$ and 2) the value of
  $V_1(T_c) =
 0.5$ GeV
 taken from lattice data \cite{19} (and quantitatively close to
 the value from the
 analytic form \cite{14}).

 We take $G_2$ in the limits 0.004 GeV$^4 \leq G_2\leq0.015$
 GeV$^4$, the value  $G_2=0.008 $GeV$^4$ ($\Delta G_2=0.0034 $ GeV$^4$)  being
in
 agreement with lattice data of $T_c(0)$, for
 $n_f=0,2,3$, see Table 1.\\

\vspace{0.5cm}

Table 1. The values of  $T_c(\mu=0)$ and $\mu_c(T=0)$ computed
using (\ref{14}) and (\ref{15}) for several values of  $\Delta
G_2$ and $ n_f=0,2,3$ \vspace{0.5cm}

\begin{center}

 \begin{tabular}{|l|l|l|l|l|}\hline
   &&&&\\
$\frac{\Delta G_2}{0.01~{\rm  GeV}^4}$& 0.191&0.341&0.57&~~~1\\
&&&&\\
\hline
&&&&\\
 $T_c({\rm ~ GeV})$~~ $n_f=0$ &0.246&0.273&0.298&0.328\\ &&&&\\
\hline
&&&&\\
 $T_c({\rm ~ GeV})$~~ $n_f=2$ &0.168&0.19&0.21&0.236\\ &&&&\\
\hline
&&&&\\
$T_c({\rm~ GeV})$~~  $n_f=3$ &0.154&0.172&0.191&0.214\\&&&&\\
\hline
&&&&\\
$\mu_c( {\rm ~ GeV})$ ~~ $n_f=2$ & 0.576& 0.626&0.68&0.742\\&&&&\\
\hline
&&&&\\
$\mu_c( {\rm ~ GeV})$ ~~ $n_f=3$ & 0.539& 0.581&0.629&0.686\\&&&&\\
\hline

 \end{tabular}\\

\end{center}
Note that $T_c$ at $n_f=0$ in Table 1  is obtained not from
(\ref{14}), but directly from (\ref{12}) with $n_f=0, \chi(T)=0$.

 The curve in Fig.2 has the expected form, which agrees with the
 curve, obtained in \cite{22} by the reweighting technic and
  agrees  for $\mu<300$ MeV with  that, obtained by the density
 of state method \cite{26}, and by the imaginary $\mu$ method \cite{25}.

 An analysis of the integral (\ref{8}) for $\nu=0$ reveals that it
  has a mild singular point at $\mu_{sing} =\frac{V_1}{2} \pm i\pi
  T$, which may show up in derivatives in $\frac{\mu}{T}$. At
  $T=0, \mu_{sing} = \frac{V_1}{2} \cong 0.25$ GeV and is  close
  to the point where one expects irregularities on the phase
  diagram  \cite{26,28}.

The limit of small $\mu$ is given in (\ref{14}). Taking $V_1
(T_c)\approx 0.5$ GeV as follows from lattice and analytic
estimates, one obtains with $\sim 3\%$ accuracy the values of
$T_c$ given  on Fig.2 for $n_f=2,3$ and $\Delta G_2=0.0034$
GeV$^4$.

 The values of $\mu_c, $  from (\ref{15})  are given in Table 1  and are in
agreement with the curves shown in Fig.3. Note that $\chi(T=0)=0$
and (\ref{15}) holds also in the case, when hadron (and baryon)
gas is taken into account.

At this point one should stress that our calculation  in VDM of
$p(T)$ does not contain  model parameters and the only
approximation is the neglect of interparticle interaction as
compared to the interaction of each one with  the vacuum (apart
from neglect of $\chi(T)$). Fig.1 demonstrates that this
approximation is reasonably good and one expects some 10$\div15\%$
accuracy in prediction of $T_c(\mu)$. Note that in VDM the phase
transition is of the first order, which is supported for $n_f=0$
by  lattice data, see e.g. \cite{6} however for $n_f=2,3$  lattice
results disagree, see \cite{29,30} for a possible preference of
the first order transition. One however should have in mind that
the final conclusion for lattice data  depends on input  quark
masses and continuum limit.

The ``weakening'' of  the phase transition for $n_f>0$ and nonzero
quark  masses is explained in our approach by the flattening of
the curve $P(T)$ at $T\approx T_c$ when hadronic gas $\chi(T)$ is
taken into account,  since $\chi(T)$ for  $n_f=0$ is much smaller
than for $n_f=2$.

A few words about chiral symmetry properties of the transition. It
was argued in a series of papers of one of the authors, that
Chiral Symmetry Breaking (CSB) and confinement are closely
connected, and  moreover the effective scalar quark mass operator
was calculated in terms of $D^E(x) (\sigma^E)$ in \cite{31} and
chiral condensate $\lan \bar q q\ran$ and $f_\pi$ via $\sigma^E$
in \cite{32}.

Therefore $\lan \bar q q\ran$ and $f_\pi$  disappear at the same
$T_c$ where $D^E(x)$ (and $\sigma^E$) vanishes. This fact is in
perfect agreement with known lattice data and supports the results
of the present study.

Finally, as seen from our expressions for $p_q, p_g$ (\ref{4}),
(\ref{7}), our EoS is independent of $Z(N_c)$ factors and $Z(N_c)$
symmetry  is irrelevant for the NP dynamics in our approach. This
result is a consequence of more general property -- the gauge
invariance of the partition function for all $T$, $\mu$, which
requires that only closed Wilson loops appear in  the resulting
expressions, yielding finally  only absolute value of the Polyakov
loop, or in other words, is expressed via only singlet free energy
of quark and  antiquark $F_{Q\bar Q}^{(1)} \approx V_1$.

\section{Summary and conclusions}

We have calculated the QCD phase diagram in the lowest order of
the background perturbation theory taking into account the change
of gluonic vacuum condensate in the phase transition and
modification of individual quark and gluon propagators in vacuum
fields.  Some support of this independent single particle approach
can be found in the lack of strangeness $(S)$-charge  (baryonic
$B$ or electric  $Q$) correlations on the lattice \cite{33} for
$T\geq1.1 T_c$.

The resulting phase curve $T_c(\mu)$ depends only on two
fundamental parameters: 1) the change of gluonic condensate across
phase boundary, $\Delta G_2\approx \frac12 G_2$ and 2) the
absolute value of Polyakov loop at $T=T_c, L_{fund} (T_c)= \exp
\left(-\frac{V_1(T_c)}{2T_c}\right),$ or else singlet $Q\bar q$
free energy $V_1 (T_c)\cong  F^1_{Q\bar Q} (\infty, T)$.

The second quantity is known from lattice   and analytic
calculations, $V_1 (T_c) \approx 0.50(5) $ GeV.

It is shown that $T_c(\mu)$ depends on $\Delta G_2$ rather mildly
and yields reasonable values of $T_c(0) \approx 0.2$ GeV for
generally accepted $G_2\approx 0.01 $ GeV$^4$ with
$\mu_c(T=0)\approx 0.6$ GeV. Moreover for the same value of
$\Delta G_2=0.00341$ GeV$^4$ one obtains the set of three values
$T_c(0) =(0.27; 0.19; 0.17)$ GeV$^4$ all being in agreement with
lattice data.

Our  results do not contain model or fitting parameters and allow
in this  (however crude) approximation connect in a very simple
way the  phase boundary to the fundamental properties of the QCD
vacuum.

Many additional quantities like quark number susceptibilities,
baryon -strangeness correlation  etc. can be easily calculated
in the method and shall  be given elsewhere, as well as influence
of possible diquark pairing.

The authors are grateful for discussions to N.O.Agasian,
K.G.Boreskov, B.L.Ioffe, A.B.Kaidalov, O.V.Kancheli, B.O.Kerbikov.

The financial support of RFBR grants 06-02-17012,  grant for
scientific schools NSh-843.2006.2 and the State Contract
02.445.11.7424 is gratefully acknowledged. This work was done with
financial support of the Federal Agency of Atomic Energy.

\end{document}